\documentclass[conference]{IEEEtran}
\IEEEoverridecommandlockouts
\usepackage{cite}
\usepackage{url}
\usepackage{amsmath,amssymb,amsfonts}
\usepackage{graphicx}
\usepackage{hyperref}
\usepackage{textcomp}
\usepackage{enumitem}
\usepackage{filecontents}
\usepackage{soul}
\usepackage{xspace}
\usepackage{tikz}
\usepackage{minted}
\usepackage{xcolor}
\usepackage{pifont}
\usepackage{array}
\usepackage{makecell}
\usepackage{multirow}
\usepackage{tabularx}
\usepackage{booktabs}   
\usepackage{subcaption} 
\usepackage{algpseudocode}
\usepackage[linesnumbered, ruled,vlined]{algorithm2e}

\newcommand{\SysName}{\texttt{OctoPipe}}

\newcommand*\emptycirc[1][1ex]{\raisebox{-0.4ex}{\tikz\draw (0,0) circle (#1);}}

\newcommand*\fullcirc[1][1ex]{\raisebox{-0.4ex}{\tikz\fill (0,0) circle (#1);}}
\newcommand\wcircle[1]{\raisebox{-0.2ex}{\Large\ding{\numexpr171+#1}}}

\def\BibTeX{{\rm B\kern-.05em{\sc i\kern-.025em b}\kern-.08em
    T\kern-.1667em\lower.7ex\hbox{E}\kern-.125emX}}
\begin{document}

\title{\Large \bf \SysName{}: Reducing Pipeline Bubbles for Heterogeneous Models via Co-Optimizing Partitioning, Placement, and Scheduling}

\author{
\begin{tabular}{@{}ccc@{}}
\begin{minipage}[t]{0.31\textwidth}\centering
Jihu Guo\textsuperscript{*}\\
\textit{FDU \& Shanghai AI Laboratory}\\
jhguo24@m.fudan.edu.cn
\end{minipage}
&
\begin{minipage}[t]{0.31\textwidth}\centering
Tenghui Ma\textsuperscript{*}\\
\textit{FDU \& Shanghai AI Laboratory}\\
thma24@m.fudan.edu.cn
\end{minipage}
&
\begin{minipage}[t]{0.31\textwidth}\centering
Wei Gao\textsuperscript{\dag}\\
\textit{HKUST}\\
csgaowei@ust.hk
\end{minipage}
\\[7ex]
\begin{minipage}[t]{0.31\textwidth}\centering
Peng Sun\\
\textit{Independent Researcher}\\
sunpengsdu@gmail.com
\end{minipage}
&
\begin{minipage}[t]{0.31\textwidth}\centering
Xun Chen\\
\textit{Independent Researcher}\\
buaa7590@gmail.com
\end{minipage}
&
\begin{minipage}[t]{0.31\textwidth}\centering
Jiaxing Li\\
\textit{Independent Researcher}\\
li126com2@gmail.com
\end{minipage}
\\[7ex]
\begin{minipage}[t]{0.31\textwidth}\centering
Zhisheng Ye\\
\textit{Independent Researcher}\\
yezhisheng@pku.edu.cn
\end{minipage}
&
\begin{minipage}[t]{0.31\textwidth}\centering
Yuyang Jin\textsuperscript{\dag}\\
\textit{Tsinghua University}\\
jinyuyang@tsinghua.edu.cn
\end{minipage}
&
\begin{minipage}[t]{0.31\textwidth}\centering
Dahua Lin\\
\textit{CUHK \& SenseTime Research}\\
dhlin@ie.cuhk.edu.hk
\end{minipage}
\end{tabular}
\thanks{\textsuperscript{*}Equal contribution. \textsuperscript{\dag}Corresponding author.}
}

\maketitle

\begin{abstract}
Pipeline parallelism is widely used to train large language models (LLMs). However, increasing heterogeneity in model architectures exacerbates pipeline bubbles, thereby reducing training efficiency. Prior approaches typically optimize a single phase of the pipeline schedule (i.e., partitioning, placement, or scheduling), leaving substantial pipeline bubbles. While promising, co-optimization poses three key challenges: (1) complex performance modeling, (2) a combinatorial search space, and (3) irregular execution orders.
To address these challenges, we propose \SysName{}, a pipeline parallelism system to jointly optimize partitioning, placement, and scheduling. First, we build a graph-based pipeline simulator to model heterogeneous pipeline execution for co-optimization. Second, on top of the simulator, we develop an iterative bubble-aware tuner to efficiently explore the combinatorial search space. Third, we implement a unified pipeline executor that dynamically orchestrates computation and communication to support irregular execution orders without deadlocks while maximizing communication-computation overlap.
Experiments show that \SysName{} achieves 1.09--1.49$\times$ throughput improvement over the state-of-the-art pipeline parallelism approaches across various heterogeneous model configurations and GPU cluster scales.
\end{abstract}

\section{Introduction}

Training large language models (LLMs) at scale requires distributing computation across a large number of accelerators~\cite{MegaScale,narayanan2021megatron2interleaved1F1B,huang2019gpipe}. Pipeline parallelism (PP)~\cite{huang2019gpipe} is widely adopted to enable such scaling~\cite{liu2024deepseekv3, dubey2024llama3, kimik2, gptoss, yang2025qwen3}. The efficiency of PP depends on three coupled design phases:
\textit{(1) Partitioning}: dividing the model into a sequence of stages, each consisting of consecutive model layers.
\textit{(2) Placement}: mapping stages to devices such as GPUs.
\textit{(3) Scheduling}: determining the execution order of micro-batches across stages for forward and backward passes.
Each phase admits multiple policies, and different combinations of these policies yield distinct pipeline schedules (e.g., AFAB~\cite{huang2019gpipe}, 1F1B~\cite{fan2021dapple}). A pipeline executor then runs the training process according to a chosen pipeline schedule~\cite{huang2019gpipe, narayanan2019pipedream, fan2021dapple,narayanan2021megatron2interleaved1F1B,jiang2024dynapipe,liu2024deepseekv3}.  Yet the inter-stage dependencies in PP inevitably create pipeline bubbles (i.e., device idle time), which harm training efficiency~\cite{huang2019gpipe, narayanan2021megatron2interleaved1F1B,liu2024deepseekv3,fan2021dapple}.

This inefficiency becomes more pronounced when training heterogeneous models. Homogeneous models consist of repeated identical layers (e.g., LLaMA~\cite{touvron2023llama2, dubey2024llama3} and GPT~\cite{radford2019gpt2, brown2020gpt3} series), whereas heterogeneous models incorporate multiple layer types~\cite{liu2024deepseekv3,nemotronnano2,kimik2,li2025minimax}. For example, Jamba~\cite{lieber2024jamba} interleaves sparse Mixture-of-Experts (MoE)~\cite{lepikhin2020gshard} layers with dense Feed-Forward Network (FFN) layers, while Nemotron~\cite{blakeman2025nemotronh,nemotronnano2} combines Mamba~\cite{gu2023mamba} and Softmax Attention~\cite{vaswani2017attention} layers. Because these layer types have markedly different computation and memory characteristics, they often create substantial stage imbalance in pipeline execution~\cite{yeung2024vocpara,zhu2025mist,zheng2022alpa}. As a result, we find that the widely used 1F1B schedule~\cite{fan2021dapple} incurs significantly higher bubble ratios on heterogeneous models than on homogeneous ones (\S\ref{sec:trendofheter}).

To reduce pipeline bubbles, prior PP approaches~\cite{zhu2025mist, sun2024adapipe, um2024metis, zheng2022alpa, fan2021dapple, narayanan2019pipedream, liu2024deepseekv3, lamy2023breadth, liu2023hanayo, narayanan2021megatron2interleaved1F1B, li2021chimera, lin2024tessel, jiang2024dynapipe, qi2024zerobubble} primarily focus on optimizing a single phase of the pipeline. 
One line of work~\cite{sun2024adapipe, zhu2025mist, um2024metis, zheng2022alpa, fan2021dapple} improves the partitioning by adjusting the number of layers per stage to alleviate computation imbalance across devices. Another line of work~\cite{lamy2023breadth, liu2023hanayo, li2021chimera, narayanan2021megatron2interleaved1F1B} tunes the placement either by splitting stages into smaller virtual stages~\cite{lamy2023breadth, liu2023hanayo, narayanan2021megatron2interleaved1F1B} or by replicating stages~\cite{liu2024deepseekv3,li2021chimera} to improve device utilization. A third line of work~\cite{lin2024tessel,jiang2024dynapipe,qi2024zerobubble} optimizes scheduling by reordering the execution of forward and backward passes to fill bubbles. While optimizing any single phase can reduce bubbles, the interactions among the remaining phases are left unaddressed, leading to suboptimal pipeline schedules. This observation motivates \emph{co-optimizing partitioning, placement, and scheduling} to further improve training efficiency (\S\ref{sec:mot_coopt}).

Nevertheless, co-optimizing these phases for heterogeneous models introduces three primary challenges.
\textbf{(C1) Complex Pipeline Performance Modeling}: Joint optimization creates intricate cross-phase interactions, and heterogeneous layers with diverse computation and memory characteristics further complicate accurate performance estimation. As a result, closed-form analytical models struggle to capture end-to-end latency and time-varying memory usage with sufficient fidelity~\cite{zheng2022alpa,um2024metis,zhu2025mist}. \textbf{(C2) Combinatorial Search Space}: Co-optimizing three phases yields an exponentially large design space with respect to the number of model layers, stages, and micro-batches, making exhaustive exploration impractical.
\textbf{(C3) Irregular Execution Orders}: The resulting pipeline schedules can induce irregular orders of computation and communication, violating the regular execution patterns assumed by prior executors~\cite{fan2021dapple,qi2024zerobubble,jiang2024dynapipe}. Without careful coordination, such irregular execution orders can trigger communication deadlocks or reintroduce pipeline bubbles (\S\ref{sec:mot_challenge}).

To address these challenges, we propose \SysName{}, a pipeline parallelism system that enables accurate modeling, efficient co-optimization, and reliable execution of irregular pipeline schedules. 
To address \textbf{C1}, we develop a graph-based pipeline simulator (\S\ref{sec:performance_model}). 
The simulator first profiles the computation, communication, and memory costs of heterogeneous layers offline. Given a candidate pipeline schedule, it constructs a Directed Acyclic Graph (DAG) that captures the dependencies and interactions among computation and communication events under resource constraints. By traversing this DAG, the simulator produces fine-grained timelines of computation, communication, and memory usage, enabling accurate resource tracing to guide subsequent co-optimization.

To address \textbf{C2}, we develop an iterative, bubble-aware pipeline tuner (\S\ref{sec:pipeline_tuner}). The tuner explores the joint design space of partitioning, placement, and scheduling. In each iteration, it applies phase-specific heuristics with the following priorities: (1) partitioning to balance workloads across devices, (2) placement to reduce boundary bubbles, and (3) scheduling to eliminate residual bubbles. Throughout the search, our graph-based pipeline simulator provides bubble breakdowns and GPU memory traces, enabling the tuner to target the dominant bottleneck while avoiding pipeline schedules that may trigger out-of-memory (OOM) errors. This iterative strategy decomposes the joint optimization into a sequence of phase-level bubble reductions, substantially shrinking the effective search space and producing efficient pipeline schedules (\autoref{fig:convergency_nemotronh}).

To address \textbf{C3}, we implement a unified pipeline executor (\S\ref{sec:pipeline_executor}). Because our tuner can generate irregular orders of computation and communication, existing executors~\cite{fan2021dapple,qi2024zerobubble,jiang2024dynapipe}, which are designed around specific phase abstractions and fixed execution patterns, cannot support such schedules directly (\S\ref{sec:2_executor}). Our executor instead adopts a DAG-driven policy that derives execution dependencies directly from the pipeline DAG. It further applies a deadlock-free, overlap-aware reordering policy to safely coordinate computation and communication to prevent communication deadlocks while maximizing communication-computation overlap.

We implement \SysName{} on top of Megatron-LM~\cite{shoeybi2019megatron}. 
\SysName{} achieves an average prediction error of 3.38\% for execution time and 5.53\% for peak memory and enables efficient pipeline schedule search within the combinatorial co-optimization space. 
Across diverse heterogeneous models and cluster scales up to 128 NVIDIA H800 GPUs, \SysName{} delivers 1.09--1.49$\times$ training throughput improvement over four state-of-the-art pipeline parallelism systems.

\section{Background}
\subsection{Parallelisms in Distributed Training}
\label{subsec:dist_para}

LLM distributed training typically requires combining multiple parallelisms to achieve scalable training across large GPU clusters~\cite{narayanan2021megatron2interleaved1F1B, liu2024deepseekv3, zhu2025mist, zheng2022alpa, MegaScale}. 
\noindent\textbf{Data Parallelism (DP)}~\cite{li2020pytorchddp, sergeev2018horovod} splits the input training data batch into smaller mini-batches. Each device maintains a copy of model parameters and processes its assigned mini-batches. 
\noindent\textbf{Tensor Parallelism (TP)}~\cite{shoeybi2019megatron} partitions a layer into smaller layer shards across multiple GPUs. TP alleviates GPU memory pressure but requires intensive collective communication (e.g., \texttt{all-gather} and \texttt{reduce-scatter}). 

\begin{figure}
    \centering
    \includegraphics[width=1\linewidth]{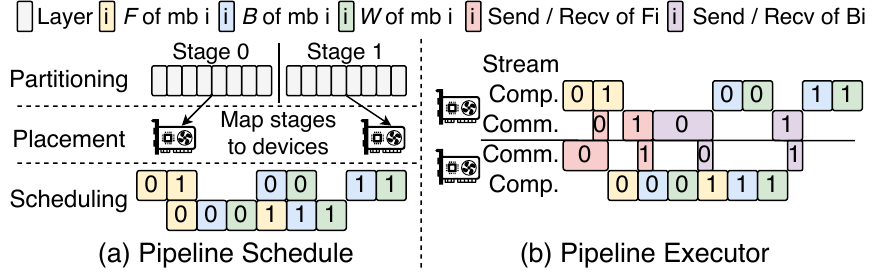}
    \caption{(a) Illustration of three pipeline phases: Partitioning, Placement, and Scheduling in a Pipeline Schedule. (b) Illustration of executing a pipeline schedule with a Pipeline Executor.}
    \label{fig:pipeline_phases}
    \vspace{-15pt}
\end{figure}

\textbf{Pipeline Parallelism (PP)}~\cite{huang2019gpipe} typically involves three design phases: partitioning, placement, and scheduling (\autoref{fig:pipeline_phases}(a)). 
\textit{(1) The partitioning phase} determines how the model is divided into stages, each consisting of consecutive model layers. 
\textit{(2) The placement phase} determines how these stages are mapped to devices, thereby establishing inter-device data dependencies. Intermediate results are transmitted via point-to-point (P2P) communication between devices hosting dependent stages.
\textit{(3) The scheduling phase} determines the execution order of micro-batches across stages for forward and backward passes. 
Each phase admits multiple policies, and different combinations of these policies lead to different pipeline schedules. 
A pipeline executor then runs the training process according to the chosen pipeline schedule. 
We next introduce these three phases and the pipeline executor.

\subsection{Pipeline Phases}
\noindent\textbf{Partitioning}.
A common partitioning strategy is to evenly divide model layers across stages, with the input layer assigned to the first stage and the output layer to the final stage~\cite{liu2023hanayo, fan2021dapple, narayanan2021megatron2interleaved1F1B, qi2024zerobubble, jiang2024dynapipe, PPipe}. Recent approaches adjust the number of layers per stage~\cite{zhu2025mist, zheng2022alpa, um2024metis, sun2024adapipe}. These approaches formulate the partitioning task as an Integer Linear Programming (ILP) problem, which is solved through dynamic programming~\cite{sun2024adapipe, um2024metis} or ILP solvers~\cite{gurobi, de2008z3}. While these approaches mitigate imbalance to some extent, bubbles remain substantial while training heterogeneous models (\autoref{fig:3_co_opt}).

\noindent\textbf{Placement}.
A widely adopted strategy is to sequentially map the stages to devices~\cite{fan2021dapple, qi2024zerobubble, narayanan2019pipedream, sun2024adapipe, jiang2024dynapipe, zheng2022alpa, um2024metis}, assuming the number of stages equals the PP size. Interleaved~\cite{narayanan2021megatron2interleaved1F1B} introduces virtual stages, which split stages into smaller ones, to reduce bubbles at the beginning and end of the pipeline (i.e., boundary bubbles). Hanayo~\cite{liu2023hanayo} builds upon this idea but applies a wave-like stage-to-device mapping strategy. Chimera~\cite{li2021chimera} duplicates stages to form multiple pipelines, enabling concurrent execution of multiple micro-batches, at the cost of additional memory overhead. However, these placement strategies primarily target homogeneous models. On heterogeneous models, imbalanced computation across stages may introduce bubbles.

\noindent\textbf{Scheduling.}
Scheduling determines the execution order of micro-batches. With \textit{backward splitting}~\cite{oh2022ooo, qi2024zerobubble, liu2024deepseekv3, ReCycle}, the backward pass is further decomposed into input gradient computation ($B$) and parameter gradient computation ($W$), enlarging the scheduling space.
Existing scheduling policies can be broadly categorized into two paradigms. 
The first adopts fixed scheduling policies (e.g., GPipe~\cite{huang2019gpipe}, DualPipe~\cite{liu2024deepseekv3}, and Interleaved~\cite{narayanan2021megatron2interleaved1F1B}), which rely on predefined execution orders. While efficient, these approaches lack the flexibility to adapt to heterogeneous models.
The second formulates scheduling as an optimization problem and uses ILP solvers to derive execution orders (e.g., ZeroBubble~\cite{qi2024zerobubble}, DynaPipe~\cite{jiang2024dynapipe}, and Tessel~\cite{lin2024tessel}). Although more flexible, these approaches suffer from poor scalability due to the rapidly growing search space with increasing numbers of layers, stages, and micro-batches (\autoref{fig:convergency_nemotronh}). Furthermore, they typically assume fixed partitioning and placement, preventing joint optimization.

\begin{table}[t]
\caption{Taxonomy of existing pipeline parallelism approaches. \fullcirc{} : support, \emptycirc{} : not support.}
\vspace{-5pt}
\centering
\footnotesize
\begin{tabular}{l@{\hspace{-4pt}}c@{\hspace{6pt}}c@{\hspace{6pt}}c|c}
\toprule
&\makecell{Partitioning\\Tuning}
&\makecell{Placement\\Tuning}
&\makecell{Scheduling\\Tuning}
&\makecell{Joint Opt.} \\\midrule
Mist~\cite{zhu2025mist} &\fullcirc &\emptycirc &\emptycirc &\emptycirc\\
Alpa~\cite{zheng2022alpa} &\fullcirc &\emptycirc &\emptycirc &\emptycirc\\
Dapple~\cite{fan2021dapple}      &\fullcirc &\emptycirc &\emptycirc &\emptycirc\\
AdaPipe~\cite{sun2024adapipe} &\fullcirc &\emptycirc &\emptycirc &\emptycirc\\
Metis~\cite{um2024metis} &\fullcirc &\emptycirc &\emptycirc &\emptycirc\\
Interleaved~\cite{narayanan2021megatron2interleaved1F1B} &\emptycirc &\fullcirc &\emptycirc &\emptycirc\\
DualPipe~\cite{liu2024deepseekv3} &\emptycirc &\fullcirc &\emptycirc &\emptycirc\\
Hanayo~\cite{liu2023hanayo} &\emptycirc &\fullcirc &\emptycirc &\emptycirc\\
Chimera~\cite{li2021chimera} &\emptycirc &\fullcirc &\emptycirc &\emptycirc\\
Tessel~\cite{lin2024tessel} &\emptycirc &\emptycirc &\fullcirc &\emptycirc\\
DynaPipe~\cite{jiang2024dynapipe} &\emptycirc &\emptycirc &\fullcirc &\emptycirc\\
ZeroBubble~\cite{qi2024zerobubble}         &\emptycirc &\emptycirc &\fullcirc &\emptycirc\\
Mario~\cite{liu2025mario} &\emptycirc &\emptycirc &\fullcirc &\emptycirc\\
\SysName{} (Ours) &\fullcirc &\fullcirc &\fullcirc &\fullcirc\\
\bottomrule
\end{tabular}
\label{tab:existing_methods}
\vspace{-10pt}
\end{table}
\subsection{Pipeline Executor}\label{sec:2_executor}
The pipeline executor orchestrates computation and communication on each device to materialize a pipeline schedule (\autoref{fig:pipeline_phases}(b)). Prior executor designs tightly couple execution logic with specific phase policies~\cite{huang2019gpipe, fan2021dapple, narayanan2021megatron2interleaved1F1B, jiang2024dynapipe}, assuming fixed and regular execution orders. Specifically, the 1F1B executor enforces a fixed execution order that repeatedly performs receive, forward/backward, and send operations on each device~\cite{megatron-lm-github}, tailored to the 1F1B pipeline schedule~\cite{fan2021dapple}. 

However, this tight coupling limits their ability to adapt to irregular execution orders and prevents them from directly supporting new phase designs without modifying the executor logic. For example, Interleaved~\cite{narayanan2021megatron2interleaved1F1B} modifies the execution order of 1F1B to support virtual stages, while DynaPipe~\cite{jiang2024dynapipe} implements multiple executors to support different scheduling policies. As a result, such executors cannot natively support co-optimized pipelines, which induce irregular execution orders that violate the phase-specific execution order assumptions in existing executors.
\section{Motivation}\label{sec:limitation-of-existing-pp}

\begin{figure}
    \centering
    \includegraphics[width=1\linewidth]{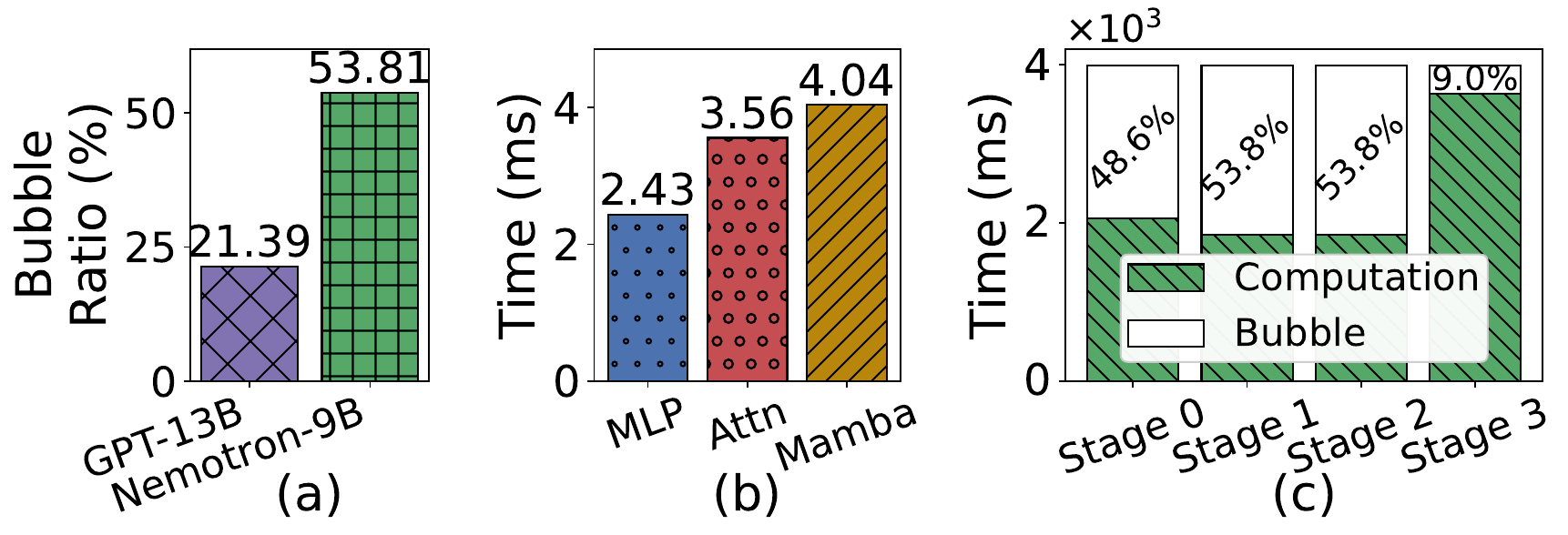}
    \vspace{-15pt}
    \caption{\textbf{[Motivation]} (a) Nemotron-9B~\cite{nemotronnano2} shows a higher bubble ratio than GPT-13B~\cite{brown2020gpt3} with 1F1B~\cite{fan2021dapple}, four stages, eight micro-batches, 4k sequence length, and four H800 GPUs. (b) Computation costs of different layer types in Nemotron-9B. (c) Imbalanced computation across stages in Nemotron-9B. }
    \label{fig:combined_1f1b_heter_bubble}
    \vspace{-15pt}
\end{figure}
\subsection{Increasing Model Heterogeneity Amplifies Bubbles}\label{sec:trendofheter}
Homogeneous models~\cite{brown2020gpt3,touvron2023llama2,dubey2024llama3} are composed of a single layer type. For example, GPT~\cite{brown2020gpt3} stacks identical Transformer layers~\cite{vaswani2017attention}. In contrast, heterogeneous models~\cite{nemotronnano2,blakeman2025nemotronh,liu2024deepseekv3,gptoss,kimik2,lieber2024jamba,li2025minimax} combine multiple layer types. For instance, Nemotron~\cite{nemotronnano2} alternates Mamba~\cite{gu2023mamba}, Softmax Attention~\cite{vaswani2017attention}, and MLP layers.
We measure the bubble ratios on different models with the widely used 1F1B schedule~\cite{fan2021dapple}, four stages, eight micro-batches, and four H800 GPUs.
\autoref{fig:combined_1f1b_heter_bubble}(a) shows that the more heterogeneous Nemotron-9B~\cite{nemotronnano2} exhibits a substantially higher bubble ratio than GPT-13B~\cite{brown2020gpt3}. This arises because different layer types have different computation costs, as shown in \autoref{fig:combined_1f1b_heter_bubble}(b). In pipeline parallelism, these differences translate into significant inter-stage latency variance. \autoref{fig:combined_1f1b_heter_bubble}(c) further illustrates how this variance amplifies pipeline bubbles. Specifically, the maximum inter-stage computation-time gap is about 200\,ms in GPT~\cite{brown2020gpt3} but exceeds 1800\,ms in Nemotron~\cite{nemotronnano2}. This severe imbalance makes the slowest stage a straggler, causing faster stages to idle while waiting~\cite{understandingstragglers} and increasing the maximum bubble ratio from 21.4\% to 53.8\%. These results show that increasing model heterogeneity can substantially degrade pipeline efficiency, making bubble reduction increasingly important.

\begin{figure}[t]
    \centering
    \includegraphics[width=1\linewidth]{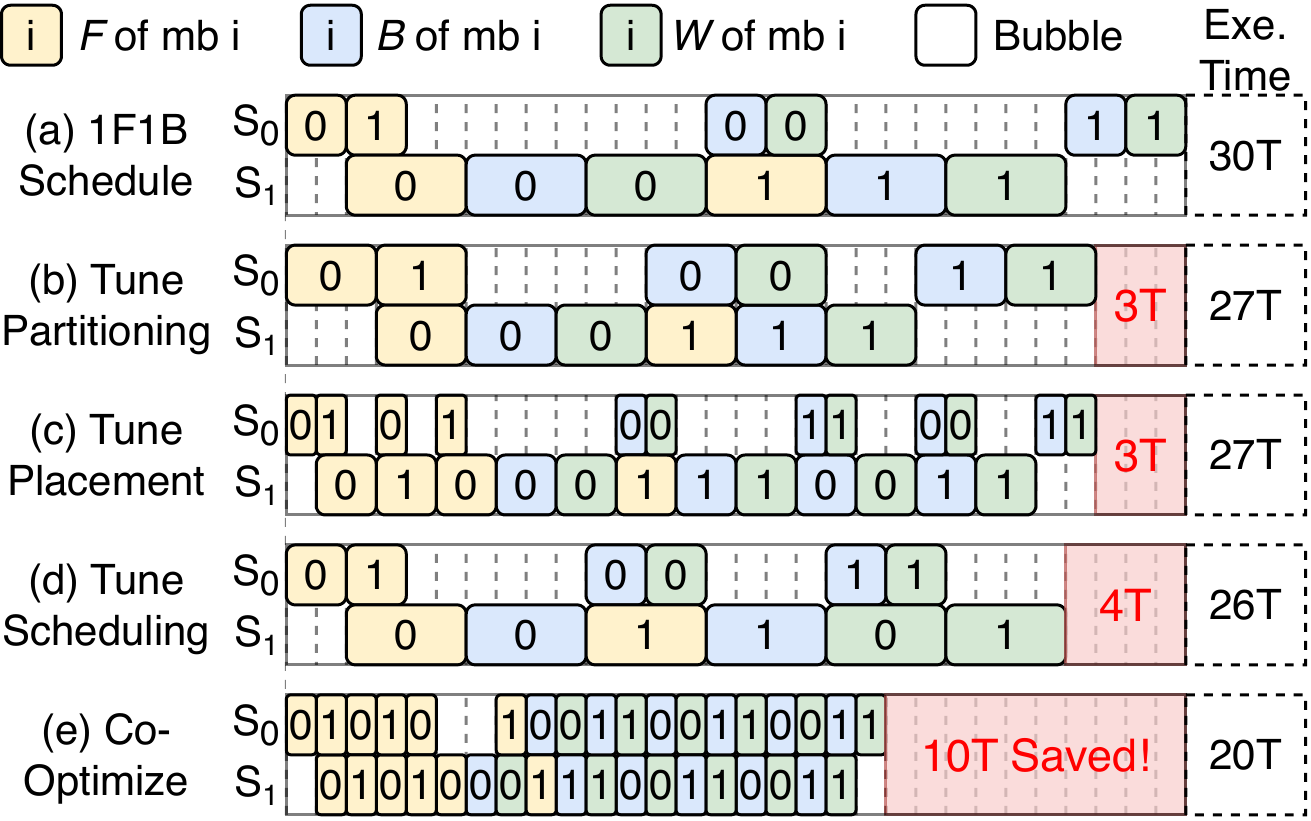}
    \vspace{-15pt}
    \caption{\textbf{[Motivation]} A brief illustration of tuning the 1F1B pipeline schedule for a heterogeneous model with two stages (S\textsubscript{0} and S\textsubscript{1}) and two micro-batches. Subfigures (b-d) show the benefit of tuning a single phase. Subfigure (e) shows the benefit of co-optimizing. $F$ denotes the forward pass, and the backward pass is split into $B$ and $W$.}
    \label{fig:3_co_opt}
    \vspace{-10pt}
\end{figure}
\subsection{Reducing Bubbles Demands Co-optimizing Pipelines}\label{sec:mot_coopt}
In \autoref{fig:3_co_opt}(a), we consider training a heterogeneous model with two stages, two micro-batches under the 1F1B pipeline schedule~\cite{shoeybi2019megatron}, where S\textsubscript{0} requires 2T for each computation workload (e.g., $F$, $B$, and $W$) and S\textsubscript{1} requires 4T due to different layer types.

\noindent\textbf{Limited Bubble Reduction of Prior Approaches.}
Prior approaches summarized in \autoref{tab:existing_methods} typically optimize only a single pipeline phase.  
As illustrated in \autoref{fig:3_co_opt}(b–d), although these approaches reduce bubbles by up to 4T, a substantial amount of bubble time remains.
Specifically, tuning partitioning balances computation workloads on S\textsubscript{0} and S\textsubscript{1} (\autoref{fig:3_co_opt}(b)), but leaves the bubbles at the beginning and end of S\textsubscript{1} unresolved. While tuning placement reduces these bubbles, the computation workloads on S\textsubscript{0} and S\textsubscript{1} remain imbalanced (\autoref{fig:3_co_opt}(c)). Tuning scheduling primarily reduces bubbles on S\textsubscript{1}, whereas bubbles on S\textsubscript{0} persist (\autoref{fig:3_co_opt}(d)). 
These examples suggest that optimizing any single phase is insufficient, leaving substantial optimization headroom unexplored (see \autoref{fig:realtraces}).

\noindent\textbf{Unlocking Optimization Headroom with Co-Optimization.}
\autoref{fig:3_co_opt}(e) presents a case study of jointly optimizing the partition, placement, and scheduling of the 1F1B~\cite{fan2021dapple} pipeline. In this example, we apply three optimizations to partitioning, placement, and scheduling, respectively: (1) moving layers from stage~1 to stage~0, (2) splitting the original two stages into six virtual stages, and (3) reordering the execution of the 1F1B pipeline.
Combining these optimizations for partition, placement, and scheduling reduces the pipeline execution time by 10T ($\approx$33\%), substantially reducing pipeline bubbles. This case study highlights the significant optimization headroom unlocked by co-optimization.

\begin{figure}
    \centering
    \includegraphics[width=1\linewidth]{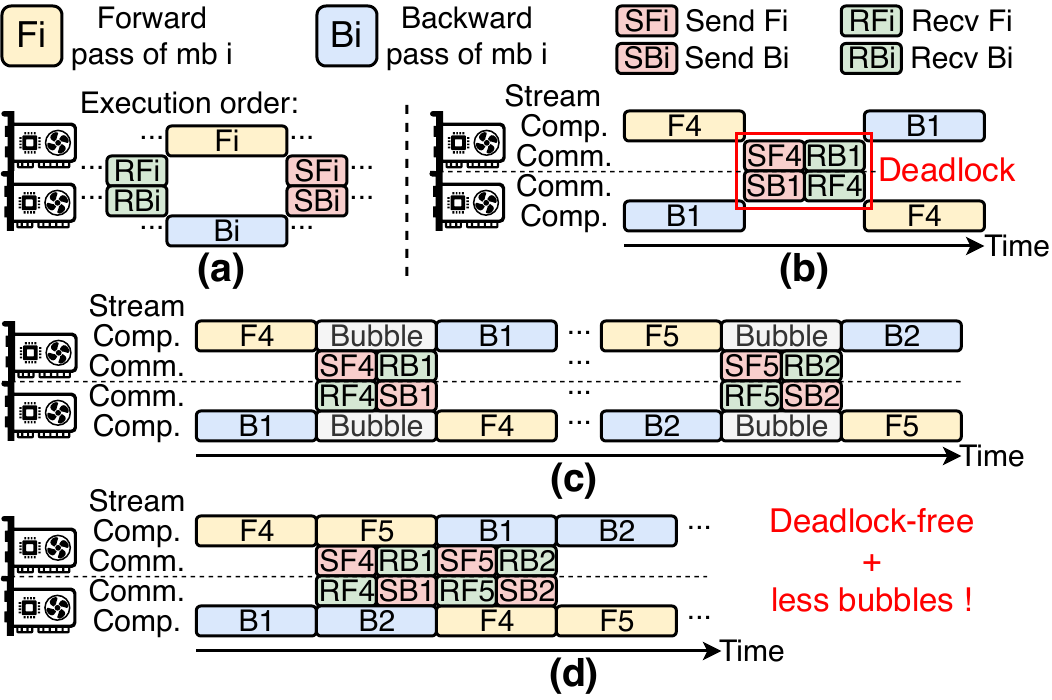}
    \caption{\textbf{[Motivation]} (a) The naive coordination of computation, send, and receive operations. (b) A deadlock caused by naive coordination. (c) Execution orders without overlap. (d) Reordering computation and communication operations to prevent deadlocks and reduce bubbles with overlap.}
    \vspace{-10pt}
    \label{fig:3_deadlock_bubble}
\end{figure}
\subsection{Co-optimizing Pipelines Remains Challenging}\label{sec:mot_challenge}
\noindent\textbf{Complex Pipeline Performance Modeling.}
Accurately evaluating pipeline performance under diverse partitioning, placement, and scheduling choices is fundamental to co-optimization. However, changing any of these phases alters the computation, communication, and memory-access patterns across GPUs. These changes affect the degree of computation--communication overlap, pipeline bubbles, and memory footprint. As a result, modeling pipeline performance with high fidelity is inherently difficult, making accurate performance estimation a major obstacle to effective co-optimization.

\noindent\textbf{Combinatorial Co-Optimization Search Space.}
Given a model with $L$ layers, $S$ stages, $N$ GPUs, and $M$ micro-batches, jointly optimizing partitioning, placement, and scheduling yields a combinatorial search space. In particular, the number of partitioning choices is $\binom{L-1}{S-1}$. Assigning $S$ stages to $N$ GPUs introduces $N^S$ placement choices. Furthermore, with $M$ micro-batches, the number of valid scheduling choices is $\frac{(2M)!}{2^M}$, since the backward pass of each micro-batch must occur after its forward pass. The resulting design space is therefore on the order of $\binom{L-1}{S-1} \times N^S \times \frac{(2M)!}{2^M}$. Such a combinatorial space makes exhaustive exploration infeasible, necessitating an efficient co-optimization algorithm.

\noindent\textbf{Irregular Execution Orders.} Co-optimization can generate irregular orders of computation and communication, violating the regular execution-order assumptions made by prior executors~\cite{fan2021dapple,qi2024zerobubble,jiang2024dynapipe}. As discussed in \S\ref{sec:2_executor}, supporting such schedules requires a more flexible executor. Moreover, without careful coordination, these irregular orders can lead to deadlocks or bubbles. The naive coordination (\autoref{fig:3_deadlock_bubble}(a)) that places a receive operation before computation and a send operation after computation can lead to deadlock when two devices concurrently issue send (or receive) operations (\autoref{fig:3_deadlock_bubble}(b))~\cite{nccl_p2p_blocking,pan2025Deadlock}. Moreover, directly executing such orders may serialize computation and communication, eliminating potential overlap and introducing bubbles (\autoref{fig:3_deadlock_bubble}(c)).
Avoiding deadlocks and reducing bubbles requires carefully coordinating computation and communication across stages (\autoref{fig:3_deadlock_bubble}(d)). However, co-optimization may generate a large number of unpredictable execution orders, making static coordination insufficient. Consequently, the executor must dynamically reorder computation and communication to avoid deadlocks and maximize overlap.

\section{\SysName{} System Overview}

\begin{figure}[t]
    \centering
    \includegraphics[width=1\linewidth]{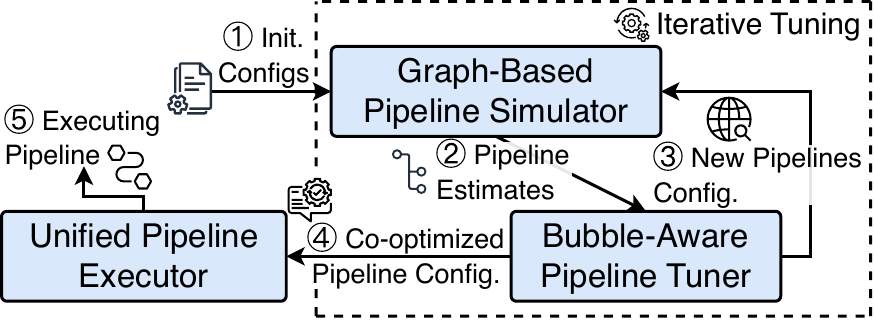}
    \vspace{-15pt}
    \caption{\SysName{} system workflow.}
    \label{fig:overview}
    \vspace{-15pt}
\end{figure}

\autoref{fig:overview} illustrates the overall workflow of \SysName{}.
\wcircle{1} \SysName{} takes as input offline profiled statistics (e.g., communication cost, per-layer computation cost, and memory usage), training configurations (e.g., \#micro-batches, \#layers, and layer types), and pipeline configurations (i.e., partition, placement, and scheduling).
\wcircle{2} The graph-based pipeline simulator (\S\ref{sec:performance_model}) constructs a DAG with these inputs and traverses the graph to simulate pipeline execution (e.g., forward, backward, send, and receive operations) and estimate performance metrics, including computation time, communication time, bubble time, and memory footprint.
\wcircle{3} Guided by these estimations, the bubble-aware pipeline tuner (\S\ref{sec:pipeline_tuner}) iteratively refines the partition, placement, and scheduling policies to explore better pipelines.
\wcircle{4} Once the search terminates, the unified pipeline executor (\S\ref{sec:pipeline_executor}) adaptively {reorders computation and communication operations} to maximize computation-communication overlap, and \wcircle{5} {executes the pipeline}.
\SysName{} operates at the PP level and is orthogonal to other parallelism schemes~\cite{lepikhin2020gshard,jacobs2023deepspeedulysses}, as well as memory-saving techniques~\cite{chen2016recomp,kirisame2020dynamicrecomp,ren2021zero-offload,chen2025SPPO,SlimPipe}.

\section{Graph-Based Pipeline Simulator}\label{sec:performance_model}

\begin{figure}
    \centering
    \includegraphics[width=1\linewidth]{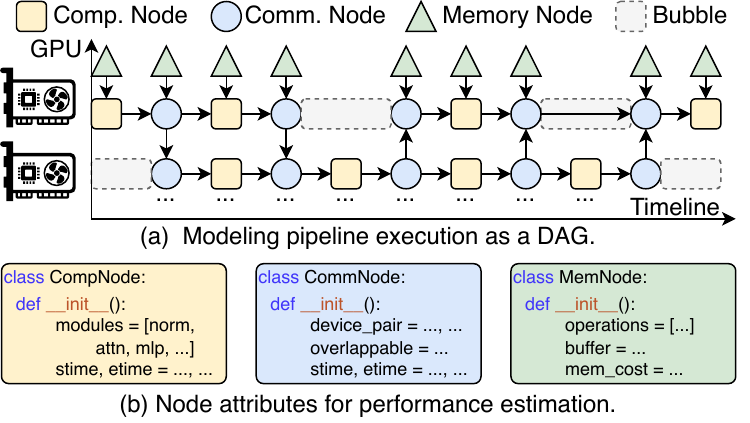}
    \vspace{-10pt}
    \caption{\SysName{} models pipeline execution as a DAG and estimates pipeline performance by traversing the graph and accumulating node costs.}
    \label{fig:5_sim}
    \vspace{-15pt}
\end{figure}

\noindent\textbf{Simulating Pipeline Execution.}
Given the input configurations, the simulator models pipeline execution as a DAG. \autoref{fig:5_sim}(a) illustrates a DAG representation of a pipeline execution with two stages and two micro-batches on two GPUs. Each DAG node represents one of three types of operations: 
(1) computation nodes for forward and backward passes of micro-batches, 
(2) communication nodes for send and receive operations, and 
(3) memory nodes for allocating and releasing intermediate tensors. 
Each computation or communication node is associated with a memory node. 
Edges in the DAG capture data dependencies between operations.

\noindent\textbf{Modeling Cross-Phase Interactions.}
The simulator updates the corresponding node attributes (\autoref{fig:5_sim}(b)) in the DAG to model cross-phase interactions when pipeline phases change. Changes to one phase (partitioning, placement, or scheduling) often propagate to the others, coupling computation, communication, and memory behaviors across phases and complicating accurate pipeline performance estimation. Such interactions are difficult to capture with closed-form analytical models~\cite{zheng2022alpa,um2024metis,zhu2025mist}.
In particular, altering partitioning changes the distribution of computation and memory across stages, placement changes the communication pattern between devices, and scheduling changes the execution order and overlap between computation and communication. To capture these effects, the simulator adjusts the \texttt{modules}, \texttt{stime} (start time), and \texttt{etime} (end time) of computation nodes and the \texttt{mem\_cost} of memory nodes to reflect partitioning changes. For placement changes, the simulator inserts or removes communication nodes corresponding to inter-device data dependencies. For scheduling changes, the simulator modifies the execution order of computation and communication nodes.
The simulator naturally extends to support memory optimization techniques~\cite{chen2016recomp,ren2021zero-offload}, which can be seamlessly modeled by adjusting the \texttt{stime} and \texttt{etime} of computation nodes and the \texttt{mem\_cost} of memory nodes.

\noindent\textbf{Estimating Pipeline Performance.}
The simulator traverses the DAG and accumulates timing and memory usage on each device to derive performance metrics. During traversal, it records the start and end time of each node and treats any idle period on a device as bubble time.
For communication nodes, the simulator determines whether send or receive operations can overlap with subsequent computation. If so, it marks them as \texttt{overlappable} and records the minimum of the computation and communication durations as the overlap time.
Using the recorded timing and memory information, the performance model derives metrics including computation time, communication time, memory usage, bubble time, and computation–communication overlap time. These metrics guide the pipeline tuner in exploring the co-optimization search space.
\section{Iterative Bubble-Aware Pipeline Tuner}\label{sec:pipeline_tuner}
\begin{figure*}[t]
    \centering
    \includegraphics[width=1\linewidth]{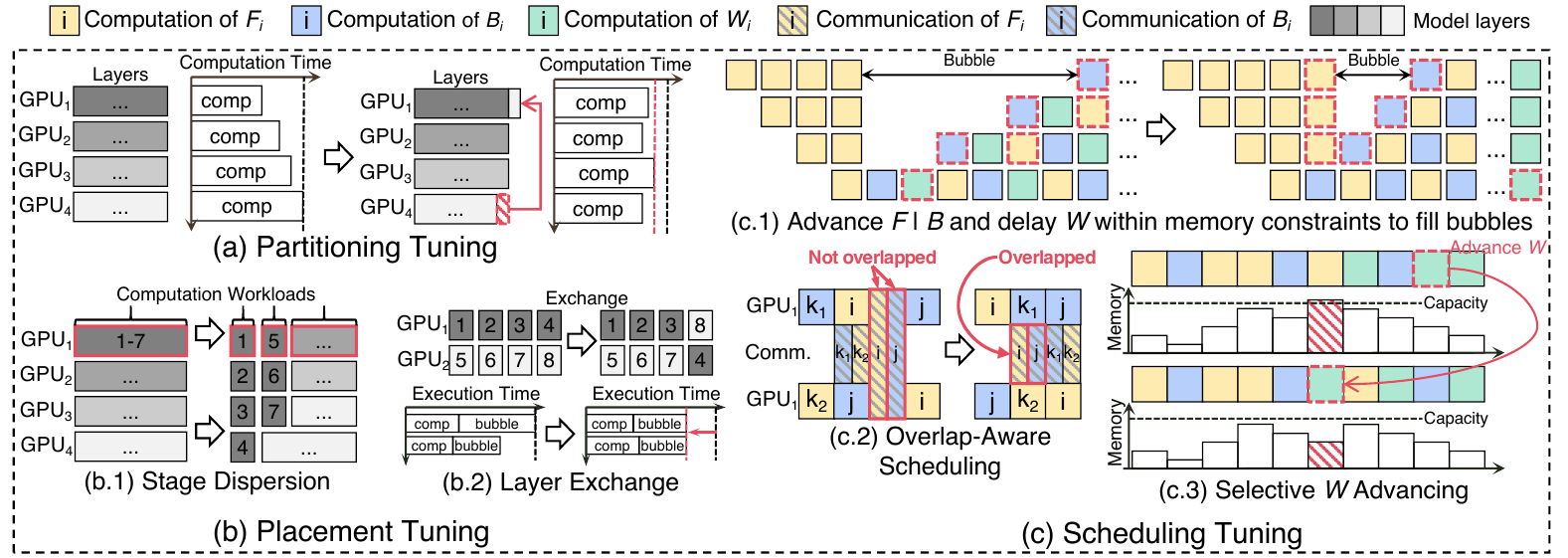}
    \vspace{-15pt}
    \caption{Illustrations of tuning (a) partitioning, (b) placement, and (c) scheduling.}
    \label{fig:pipeline_tuner}
    \vspace{-15pt}
\end{figure*}

\SysName{} employs an iterative tuning algorithm to efficiently explore the combinatorial search space of co-optimizing partitioning, placement, and scheduling. At each iteration, the tuner estimates the potential for bubble reduction in the pipeline and adjusts the phase that is most likely to reduce bubbles. The process repeats until no further performance improvement is observed for several iterations.

\noindent\textbf{Optimization Objective.}
The goal of the tuner is to minimize the pipeline execution time.
Let $\mathcal{D}$ denote the set of devices in the pipeline.
$T_d^{\text{execution}}$ denotes the execution time of device $d \in \mathcal{D}$, and the pipeline execution time is defined as
\[
T^{\text{pipeline}} = \max_{d \in \mathcal{D}} T_d^{\text{execution}} .
\]

The tuner optimizes the input partitioning, placement, and scheduling to minimize $T^{\text{pipeline}}$.
Let $L$ denote the number of layers,
$S$ the number of stages,
$R$ the number of pipeline ranks, and
$M$ the number of micro-batches.
The \textit{partitioning} is represented by a vector $\mathbf{P} \in \mathbb{Z}_{+}^{S}$, where $p_s$ denotes the number of layers assigned to stage $s$ and $\mathbf{1}^{\top}\mathbf{P} = L$.
The \textit{placement} is represented by a collection $\mathbf{A} = \{A_r \mid r \in \{1,\dots,R\}\}$, where $A_r \subseteq \{0,\dots,S-1\}$ denotes the set of stages assigned to rank $r$ and $A_r \cap A_{r'} = \emptyset \; (r \ne r'), \bigcup_{r=1}^{R} A_r = \{0,\dots,S-1\}.$
The \textit{scheduling} is represented by $\mathbf{O} = \{O_r \mid r \in \{1,\dots,R\}\}$, where $O_r \in \{fwd,bwd,send,recv\}^{M \cdot |A_r|}$ denotes the sequence of operations executed on rank $r$. Here $fwd$, $bwd$, $send$, and $recv$ represent forward, backward, send, and receive operations.
The optimization problem is formulated as
\begin{align}
\min_{\mathbf{P},\,\mathbf{A},\,\mathbf{O}}
\quad & T^{\text{pipeline}} = \max_{d \in \mathcal{D}} T_d^{\text{execution}} \\
\text{s.t.}\quad
& M_d^{\text{peak}} \le M_d^{\text{capacity}}, \quad \forall d \in \mathcal{D}.
\end{align}

\noindent\textbf{Iterative and Bubble-Aware Tuning.}
We develop an iterative, bubble-aware tuning algorithm, shown in \autoref{alg:tuning}, to efficiently explore the large co-optimization search space. In each iteration, the tuner optimizes the three phases with the following priorities: 
(1) partitioning, to balance workloads across devices;
(2) placement, to reduce boundary bubbles; and
(3) scheduling, to minimize residual bubbles.
This priority order is motivated by our empirical observation that workload imbalance is the dominant source of bubbles. Once workloads are balanced across devices, the bubbles at the pipeline boundaries account for most of the remaining bubbles. After addressing these two sources, scheduling optimization can further reduce residual bubbles.
Changing this priority order leads to slower convergence or produces worse pipelines under the same tuning time budget (see \autoref{fig:convergency_nemotronh}).
This iterative strategy decomposes the joint optimization into a sequence of phase-level bubble reductions, substantially shrinking the effective search space and producing efficient pipeline schedules.

\begin{algorithm}[t]
\small
\caption{Iterative Bubble-Aware Pipeline Tuning}
\label{alg:tuning}

\KwIn{partitioning $\mathbf{P}$, placement $\mathbf{A}$, scheduling $\mathbf{O}$, time budget $B$}

\While{tuning time $< B$}{

$\Delta{b}\leftarrow$ max bubble time gap across devices\;
$b^{\mathrm{bd}}, b^{\mathrm{res}}\leftarrow$ boundary bubble time, residual bubble time\;
$t^{\mathrm{layer}}\leftarrow$ minimum layer computation time\;

\eIf{$\Delta{b} > t^{\mathrm{layer}}$}{
    $\mathbf{P} \leftarrow \arg\min_{\mathbf{P}} T^{\mathrm{pipeline}}(\mathbf{P}, \mathbf{A}, \mathbf{O})$\;
}
{
\eIf{$b^{\mathrm{bd}} > b^{\mathrm{res}}$}{
    $\mathbf{A} \leftarrow \arg\min_{\mathbf{A}} T^{\mathrm{pipeline}}(\mathbf{P}, \mathbf{A}, \mathbf{O})$\;
}
{
    $\mathbf{O} \leftarrow \arg\min_{\mathbf{O}} T^{\mathrm{pipeline}}(\mathbf{P}, \mathbf{A}, \mathbf{O})$\;
}
}

}

\Return{$(\mathbf{P}, \mathbf{A}, \mathbf{O})$}
\end{algorithm}

\noindent\textbf{Partitioning Tuning.}
The tuner adjusts the number of layers per device to balance workloads across devices. \autoref{fig:pipeline_tuner}(a) illustrates the partitioning tuning procedure. First, the tuner fetches the workload estimate on each device from our simulator. Second, the tuner migrates layers from the most loaded device to the least loaded device to reduce imbalance. Third, the tuner modifies the DAG and calls the simulator to reevaluate the new partitioning. This migrate-and-reevaluate process repeats until no better partitioning is found.

\noindent\textbf{Placement Tuning.}
The tuner employs two policies to tune placement: stage dispersion and layer exchange. The stage dispersion refines the granularity of computation workloads. As shown in \autoref{fig:pipeline_tuner}(b.1), the tuner first identifies stages that contain multiple layers, then splits such a stage into individual layers and assigns them to different devices. The stage dispersion reduces the number of layers per stage, yields finer-grained stages, and decreases bubbles at the beginning and end of the pipeline. The layer exchange balances computation time and bubble time to minimize overall execution time. \autoref{fig:pipeline_tuner}(b.2) illustrates the layer exchange process. When $\text{GPU}_1$ and $\text{GPU}_2$ have comparable computation workloads but $\text{GPU}_1$ experiences longer bubble time, the tuner selects a layer from $\text{GPU}_1$ and a more compute-intensive layer from $\text{GPU}_2$ to swap. This exchange trades computation time for reduced bubble time, lowering the overall execution time.

While fine-grained stages reduce boundary bubbles, they introduce additional communication overhead. We mitigate this through overlap-aware scheduling (\autoref{fig:pipeline_tuner}(c.2)) and execution (\S\ref{sec:pipeline_executor}), which hide communication latency by overlapping it with computation (see \autoref{tab:comm_nemotron}).

\noindent\textbf{Scheduling Tuning}.\label{sec:scheduling_tuning}
The tuner employs three policies to tune the scheduling. First, in \autoref{fig:pipeline_tuner}(c.1), the tuner advances $F$ and $B$, and delays $W$ within the memory constraints to fill bubbles. This policy is motivated by the fact that both $F$ and $B$ exhibit inter-device data dependencies that can induce bubbles, whereas $W$ does not. For instance, the $F$ on stage $i$ depends on the result from stage $i-1$, and the $B$ on stage $i$ depends on the result from stage $i+1$. $W$ has no inter-device data dependencies since it relies only on $B$ on the same device.
This policy has two merits: (1) executing $F$ and $B$ earlier releases more $W$ in the following scheduling, and (2) the deferred $W$ can be scheduled to fill bubbles in cases where $F$ and $B$ cannot do so due to data dependency.

Second, the tuner employs an overlap-aware scheduling policy to hide communication overhead. This policy inserts other computations to overlap with communications that were previously non-overlappable due to data dependencies. In~\autoref{fig:pipeline_tuner}(c.2), the communication of $F_i$ cannot initially overlap with its own computation because of a data dependency, but $B_{k_1}$ has no dependency on the communication of $F_i$. To exploit this, the tuner inserts $B_{k_1}$ between $F_i$ and $B_j$ on GPU\textsubscript{1}. Similarly, it inserts $B_{k_2}$ between $B_j$ and $F_i$ on GPU\textsubscript{2}. After rescheduling, the communications of both $F_i$ and $B_j$ are overlapped with the computations of $B_{k_1}$ and $B_{k_2}$.

Third, the tuner leverages a selective $W$ advancing policy to prevent OOM errors. For each candidate scheduling, the tuner calls the simulator to locate memory peak usages. As shown in \autoref{fig:pipeline_tuner}(c.3), if the peak memory exceeds device capacity, the tuner advances the computation of some workload’s $W$ to lower the peak. It first searches for a $W$ whose advancement does not increase bubble time; if none exists, it advances the earliest eligible workload instead. This process repeats until the peak memory on every device fits within its capacity.

\section{Unified Pipeline Executor}\label{sec:pipeline_executor}
We design a unified pipeline executor that supports irregular execution orders while preventing deadlocks and maximizing computation-communication overlap.

\noindent\textbf{Supporting Irregular Execution Orders.}
As analyzed in \S\ref{sec:2_executor}, prior executors~\cite{fan2021dapple,narayanan2021megatron2interleaved1F1B,jiang2024dynapipe} enforce fixed computation and communication orders tailored to specific phase designs, making them ill-suited for the irregular schedules generated by co-optimization. In contrast, our executor does not rely on predefined phase-specific execution patterns, and instead derives execution orders directly from a pipeline DAG produced by our simulator (\S\ref{sec:performance_model}). The DAG consists of nodes and edges: each node represents an operation (e.g., forward, backward, send, or receive), and each edge encodes an execution dependency. Partitioning, placement, and scheduling only modify the DAG structure and node attributes, which together determine the execution order of operations on each device. Ignoring communication stalls caused by inter-device data dependencies, the executor can directly follow this derived order to execute arbitrary irregular pipeline schedules.

\noindent\textbf{Deadlock-Free and Overlap-Aware Reordering.} However, the communication stalls may cause deadlocks, discussed in \S\ref{sec:mot_challenge}. Specifically, when multiple devices simultaneously issue \texttt{send} or \texttt{recv}, they may form cyclic waiting dependencies, leading to deadlock~\cite{nccl_p2p_blocking,jiang2024dynapipe}. We propose to reorder communication operations to break such cyclic dependencies. However, reordering changes the DAG and may thereby introduce additional pipeline bubbles.

To address this issue, we design a deadlock-free, overlap-aware reordering policy that eliminates communication deadlocks while reducing pipeline bubbles. The policy consists of three steps. First, given the DAG, the executor traverses all communication operations and identifies mismatched \texttt{send}/\texttt{recv} pairs across devices that may lead to cyclic dependencies. This step takes $O(|V_c|)$ time, where $|V_c|$ is the number of communication nodes. Second, for each mismatched pair, the executor locates its dependent computation nodes and derives a feasible insertion window based on their execution intervals. This window bounds the valid positions where the communication can be reordered, effectively restricting the search space to a small local region.
Within the feasible window, the executor enumerates candidate insertion positions for the corresponding communication nodes. Let $k_i$ denote the number of candidate positions for the $i$-th communication pair; the overall search complexity is $O(\sum_i k_i)$. Third, for each candidate schedule, the executor evaluates a scheduling cost defined as the communication span, i.e., the time interval between the reordered \texttt{send} and \texttt{recv}, minus the overlap achievable with neighboring computation. The executor then greedily selects the minimum-cost candidate for each communication pair and immediately applies the reordering. Overall, this policy enables the executor to resolve deadlocks while maximizing computation-communication overlap.
\begin{figure*}[t]
    \centering
    \includegraphics[width=1\linewidth]{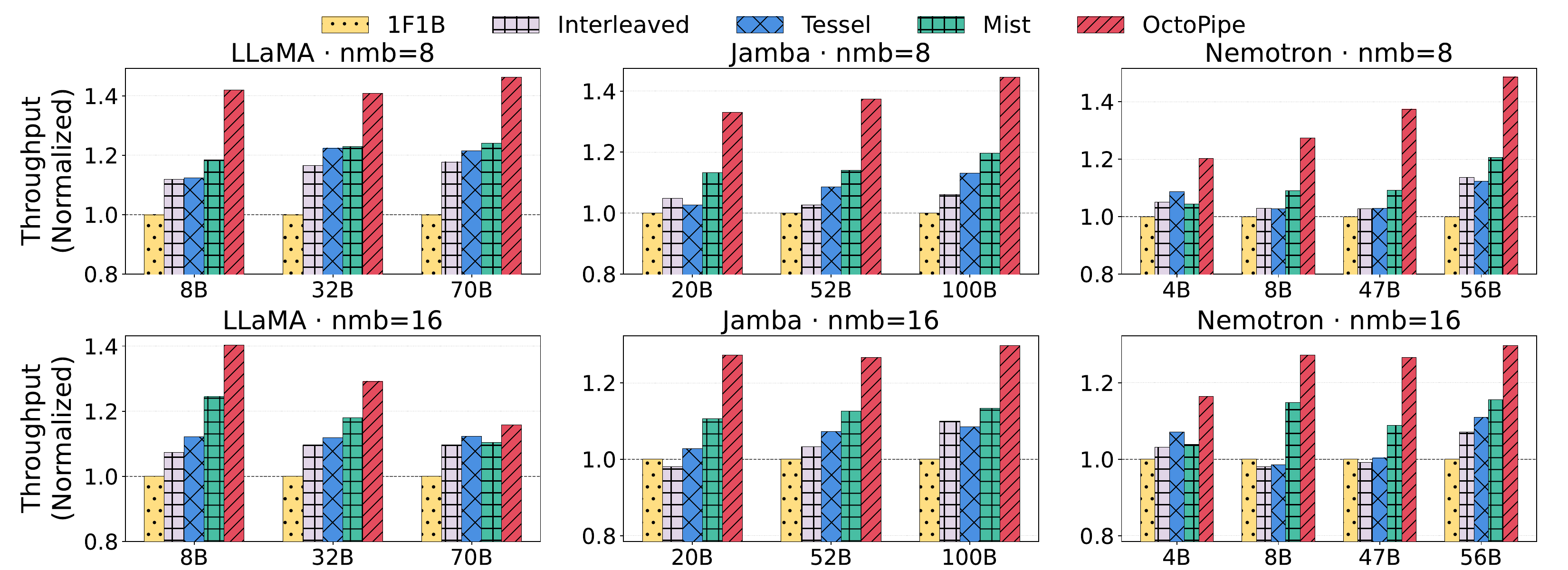}
    \vspace{-15pt}
    \caption{E2E results across homogeneous and heterogeneous models on 64 GPUs, with both PP and TP sizes set to 8.}
    \label{fig:e2etgs}
    \vspace{-15pt}
\end{figure*}
\section{Evaluation}

\noindent\textbf{Evaluation Setup.} 
We implement \SysName{} atop Megatron-LM~\cite{shoeybi2019megatron} v0.15.0 in 10k lines of Python code. We run all experiments with up to 128 NVIDIA H800 GPUs. Each node has 8 GPUs connected by 200~GB/s NVLink, and nodes are connected by an 8$\times$200~Gbps RoCEv2 network. We evaluate \SysName{} on multiple sizes of LLaMA~\cite{dubey2024llama3}, Jamba~\cite{lieber2024jamba}, and Nemotron-H~\cite{blakeman2025nemotronh} (see \autoref{tab:model_config}).

\def\modelcolhspace{5}
\begin{table}[h]
\vspace{-5pt}
\caption{Model configurations.}
\vspace{-5pt}
\centering
\footnotesize
\begin{tabular}{c@{\hspace{\modelcolhspace pt}}c@{\hspace{\modelcolhspace pt}}c@{\hspace{\modelcolhspace pt}}c@{\hspace{\modelcolhspace pt}}c@{\hspace{\modelcolhspace pt}}c@{\hspace{\modelcolhspace pt}}c}
\toprule
\textbf{Model} & \textbf{Size} & \textbf{$L$} & \textbf{$V$} & \textbf{$H$} & \textbf{FFN Type} & \textbf{Attn. Type} \\
\midrule
\multirow{3}{*}{\makecell{LLaMA\\(Homo.)}} 
 & 8B & 32  & 128k & 4096 & FFN  & SA    \\
 & 32B & 64  & 128k & 6144 & FFN  & SA    \\
 & 70B & 80 & 128k & 8192 & FFN  & SA    \\
\midrule
\multirow{3}{*}{\makecell{Jamba\\(Heter.)}}
 & 20B\footnotemark[1] & 32  & 128k & 4096 & FFN+MoE & SA+Mamba \\
 & 52B & 64  & 128k & 4096 & FFN+MoE & SA+Mamba \\
 & 100B\footnotemark[1] & 144  & 128k & 8192 & FFN+MoE & SA+Mamba \\
\midrule
\multirow{4}{*}{\makecell{Nemotron-H\\(Heter.)}}
 & 4B & 52  & 128k & 3072  & FFN  & SA+Mamba    \\
 & 8B & 52  & 128k & 4096  & FFN  & SA+Mamba    \\
 & 47B & 98  & 128k & 8192  & FFN  & SA+Mamba    \\
 & 56B & 118 & 128k & 8192  & FFN  & SA+Mamba    \\
\bottomrule
\end{tabular}
\label{tab:model_config}
\vspace{-5pt}
\end{table}

\footnotetext[1]{Jamba-20B and Jamba-100B are scaled variants of Jamba-52B and Jamba-398B, respectively, tailored to the scale of our cluster.}

\noindent\textbf{Baselines.}
We evaluate \SysName{} against the following baselines:
\textbf{(1) 1F1B}~\cite{fan2021dapple} and \textbf{(2) Interleaved}~\cite{narayanan2021megatron2interleaved1F1B}, as implemented in Megatron-LM with native communication optimizations;
\textbf{(3) Tessel}~\cite{lin2024tessel}, which adaptively reschedules workloads to minimize bubbles;
and \textbf{(4) Mist}~\cite{zhu2025mist}, an automatic training framework with adaptive partitioning based on dynamic programming.
We apply backward splitting~\cite{qi2024zerobubble} uniformly to all baselines as an orthogonal optimization to improve their throughput.
We evaluate Tessel over all predefined configurations in its implementation and report the best observed throughput.
To preserve training accuracy~\cite{narayanan2019pipedream,qi2024zerobubble}, we adopt synchronous pipeline execution for all approaches.

\noindent\textbf{Training Configurations.}
We evaluate different model sizes with PP sizes of 4, 8, and 16. The micro-batch size is fixed at 1 to reduce pipeline bubbles, while the number of micro-batches is set to either the PP size or 2$\times$ the PP size to evaluate performance under different global batch sizes. The sequence length is set to 4k. Training throughput is measured by the average Tokens/GPU/Second (TGS) over the first 100 steps, excluding the initial 10 warm-up steps.
\begin{figure}[t]
    \centering
    \includegraphics[width=1\linewidth]{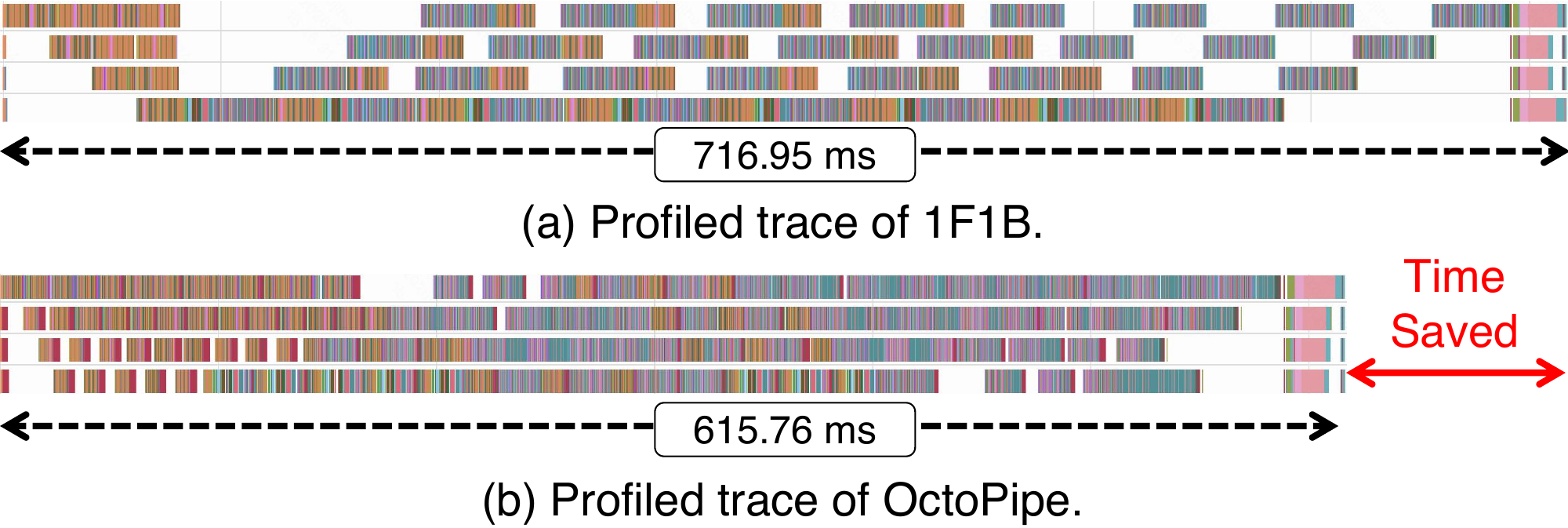}
    \vspace{-10pt}
    \caption{Traces of 1F1B and \SysName{} profiled using \texttt{torch.profiler} on Nemotron-H 8B with a PP size of 4 and eight micro-batches. White regions indicate bubbles; other regions denote GPU kernels.}
    \label{fig:realtraces}
    \vspace{-15pt}
\end{figure}

\subsection{End-to-End Evaluation}
\noindent\textbf{Throughput Results on Heterogeneous Models.}
In \autoref{fig:e2etgs}, \SysName{} consistently delivers the highest end-to-end training throughput across Jamba and Nemotron-H models, achieving 1.16$\times$--1.49$\times$, 1.13$\times$--1.36$\times$, 1.09$\times$--1.33$\times$, and 1.11$\times$--1.26$\times$ speedups over 1F1B, Interleaved, Tessel, and Mist, respectively.
These gains arise from joint optimization of partitioning, placement, and scheduling, enabled by a unified executor.
Interleaved reduces boundary bubbles by introducing virtual stages, effectively improving placement. However, it does not address stage imbalances, which are the dominant source of bubbles. Across the evaluated Jamba and Nemotron-H configurations, Interleaved ranges from a slight slowdown (0.98$\times$) to a 1.14$\times$ speedup over 1F1B.
Tessel tunes scheduling via reordering execution. However, without mitigating stage imbalance through partitioning or exposing sufficient scheduling opportunities via placement, its speedup varies from 0.99$\times$ to 1.13$\times$ over 1F1B.
Mist alleviates workload imbalance through partitioning, addressing the primary source of bubbles. However, without complementary optimizations on placement and scheduling, Mist delivers 9.8\%--20.5\% lower throughput than \SysName{}.
These results show that: (1) optimizing any single phase of the pipeline is insufficient for heterogeneous models, and (2) co-optimization effectively reduces bubbles.

\begin{figure}[t]
    \centering
    \includegraphics[width=1\linewidth]{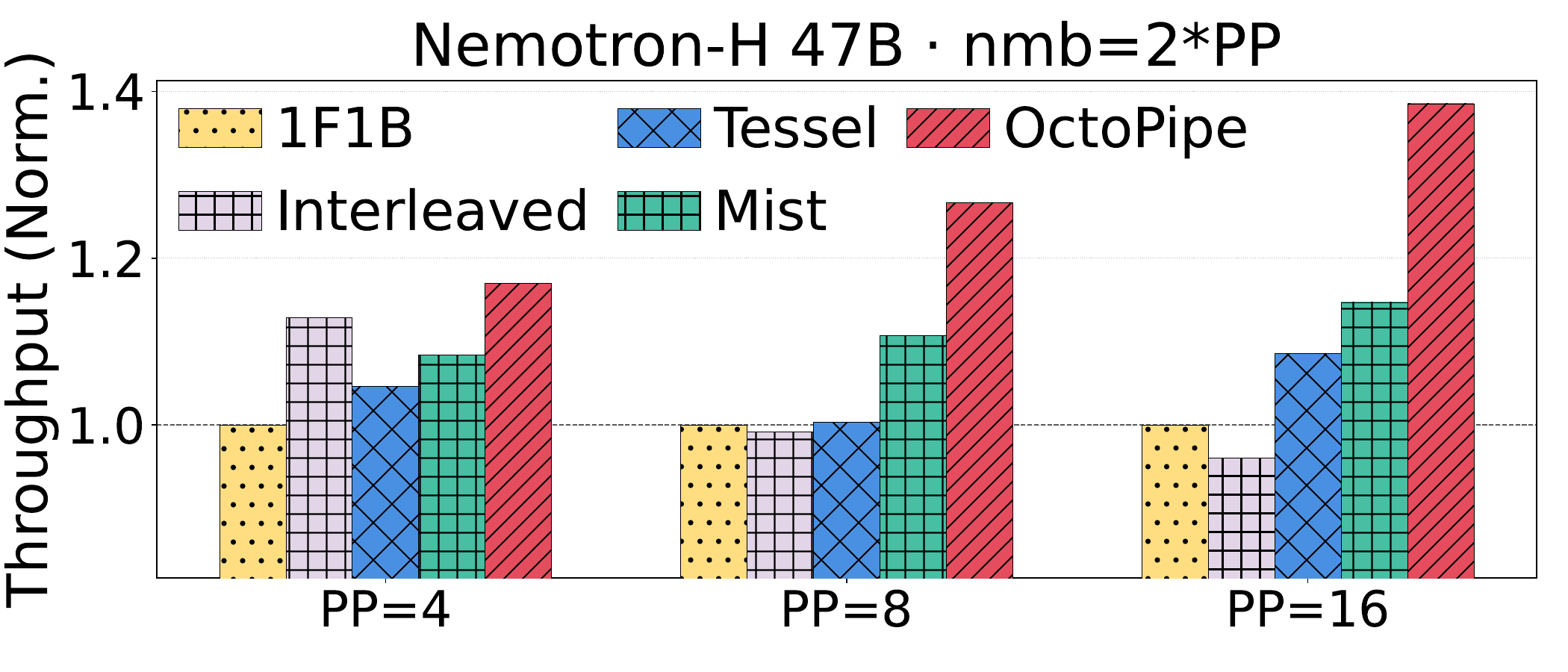}
    \vspace{-20pt}
    \caption{Throughput results across different PP sizes.}
    \label{fig:across_pp_nemotron_47b_mb_2xPP}
    \vspace{-15pt}
\end{figure}
\noindent\textbf{Impact of Model Heterogeneity on Throughput.}
The benefits of \SysName{} become more pronounced as model heterogeneity increases. On LLaMA models, where identical layers lead to well-balanced workloads across stages, the opportunity for further optimization is inherently limited. Moreover, as LLaMA size increases, the relative stage imbalance becomes smaller and the advantage of \SysName{} generally declines, especially when the number of micro-batches is set to 16. For example, the speedup over 1F1B decreases from 1.40$\times$ on 8B to 1.29$\times$ on 32B and 1.16$\times$ on 70B.
In contrast, Jamba and Nemotron-H models exhibit diverse layer types with varying computation costs, leading to pronounced stage imbalance. By jointly optimizing partitioning, placement, and scheduling, \SysName{} effectively mitigates these inefficiencies. It achieves throughput gains of 1.27$\times$--1.45$\times$ on Jamba and 1.16$\times$--1.49$\times$ on Nemotron-H, with the largest gain observed on the Nemotron-H 56B configuration.
As shown in \autoref{fig:across_pp_nemotron_47b_mb_2xPP}, the advantage of \SysName{} on Nemotron-H 47B becomes more pronounced as the PP size increases. In contrast, Interleaved can be surpassed by 1F1B at larger PP sizes because the shorter computation time of each stage makes layer-level heterogeneity account for a larger fraction of stage execution time, thereby exacerbating bubbles.

\noindent\textbf{Communication and Overlap Analysis.}
\begin{table}[t]
\vspace{5pt}
\caption{Communication statistics of Nemotron-H 56B with a PP size of 8 and eight micro-batches.}
\vspace{-10pt}
\centering
\setlength{\tabcolsep}{2pt}
\vspace{5pt}
\begin{tabular}{lccccc}
\toprule
\textbf{Method} & \textbf{\#Stages} & \textbf{\#Comm.} & \textbf{\#Overlap} & \textbf{Overlap Rate} & \textbf{Time/Step (ms)}\\
\midrule
1F1B     & 8  & 224  & 112  & 50.0\% & 2406 \\
Mist     & 8  & 224  & 112  & 50.0\% & 1996 \\
Tessel   & 16 & 480  & 320  & 66.7\% & 2142 \\
Interleaved & 56 & 1760 & 1330 & 75.6\% & 2116 \\
OctoPipe & 56 & 1760 & 1606 & 91.3\% & 1619 \\
\bottomrule
\end{tabular}
\label{tab:comm_nemotron}
\vspace{-5pt}
\end{table}
\autoref{tab:comm_nemotron} reports communication and overlap statistics for Nemotron-H 56B. Although increasing the number of stages increases total communication volume, smaller stages reduce bubbles, while effective communication--computation overlap hides much of the added communication overhead. Consequently, Tessel, Interleaved, and \SysName{} all achieve lower per-step execution time than 1F1B. These results highlight the importance of overlap-aware scheduling. \SysName{} further benefits from its executor support to realize such overlap efficiently.

\noindent\textbf{Bubble Reduction Analysis.}
\autoref{fig:realtraces} compares the profiled traces of 1F1B and \SysName{} on Nemotron-H 8B. \SysName{} significantly reduces bubbles through co-optimization, as indicated by the much smaller white regions in \autoref{fig:realtraces}(b) than in \autoref{fig:realtraces}(a). This improvement stems from (i) better workload balance across GPUs via tuning partitioning, (ii) fine-grained placement that reduces boundary bubbles, and (iii) more effective scheduling that increases computation--communication overlap and fills bubbles. These effects collectively shorten the end-to-end execution time.

Despite these improvements, marginal bubbles persist. First, a fraction of communication (approximately 8.7\% in \autoref{tab:comm_nemotron}) remains unoverlapped due to inherent data dependencies, even with our overlap-aware designs. Second, the DAG-driven executor can deviate from the original schedule under runtime variability (e.g., network jitter and hardware performance fluctuations), introducing additional inefficiencies. Nevertheless, these overheads are minor in practice and are outweighed by the gains from the co-optimization.

\begin{figure}[t]
    \centering
    \includegraphics[width=1\linewidth]{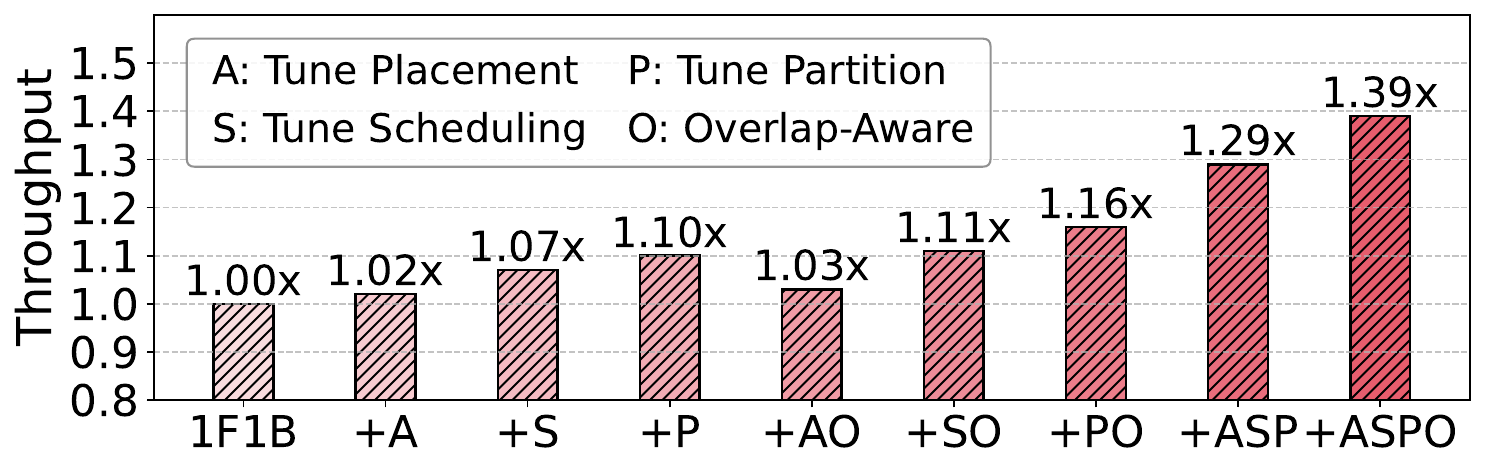}
    \vspace{-20pt}
    \caption{Performance of \SysName{} under individual optimizations and under combinations of optimizations.}
    \label{fig:ablation_study}
    \vspace{-15pt}
\end{figure}
\subsection{Performance Breakdown of \SysName{}}
\autoref{fig:ablation_study} shows the performance impact of applying individual and combined optimizations on top of 1F1B for Nemotron-H 47B with a PP size of 16 and 32 micro-batches.
Applying co-optimization (+ASP), \SysName{} achieves 1.29$\times$ training throughput over 1F1B, while incorporating the overlap-aware executor (+ASPO) further improves throughput to 1.39$\times$. In contrast, applying placement (+A), scheduling (+S), or partitioning (+P) alone yields improvements of 2\%, 7\%, and 10\%, respectively.
This result highlights that, in heterogeneous models, workload imbalance across stages is the primary source of bubbles. Partition tuning (+P), which directly mitigates such imbalances, provides the largest individual gain. However, without complementary placement and scheduling optimizations, residual bubbles remain.
The overlap-aware executor (+O) consistently improves throughput across configurations (e.g., +AO, +SO, +PO), and delivers the largest gain when combined with co-optimization (+ASPO). This is because it explores execution orders that maximize communication-computation overlap, while co-optimization exposes more opportunities for such overlap.
Overall, these results demonstrate that addressing bubbles requires both co-optimizing and exploiting overlap via the executor.

\begin{figure}[t]
    \centering
    \includegraphics[width=1\linewidth]{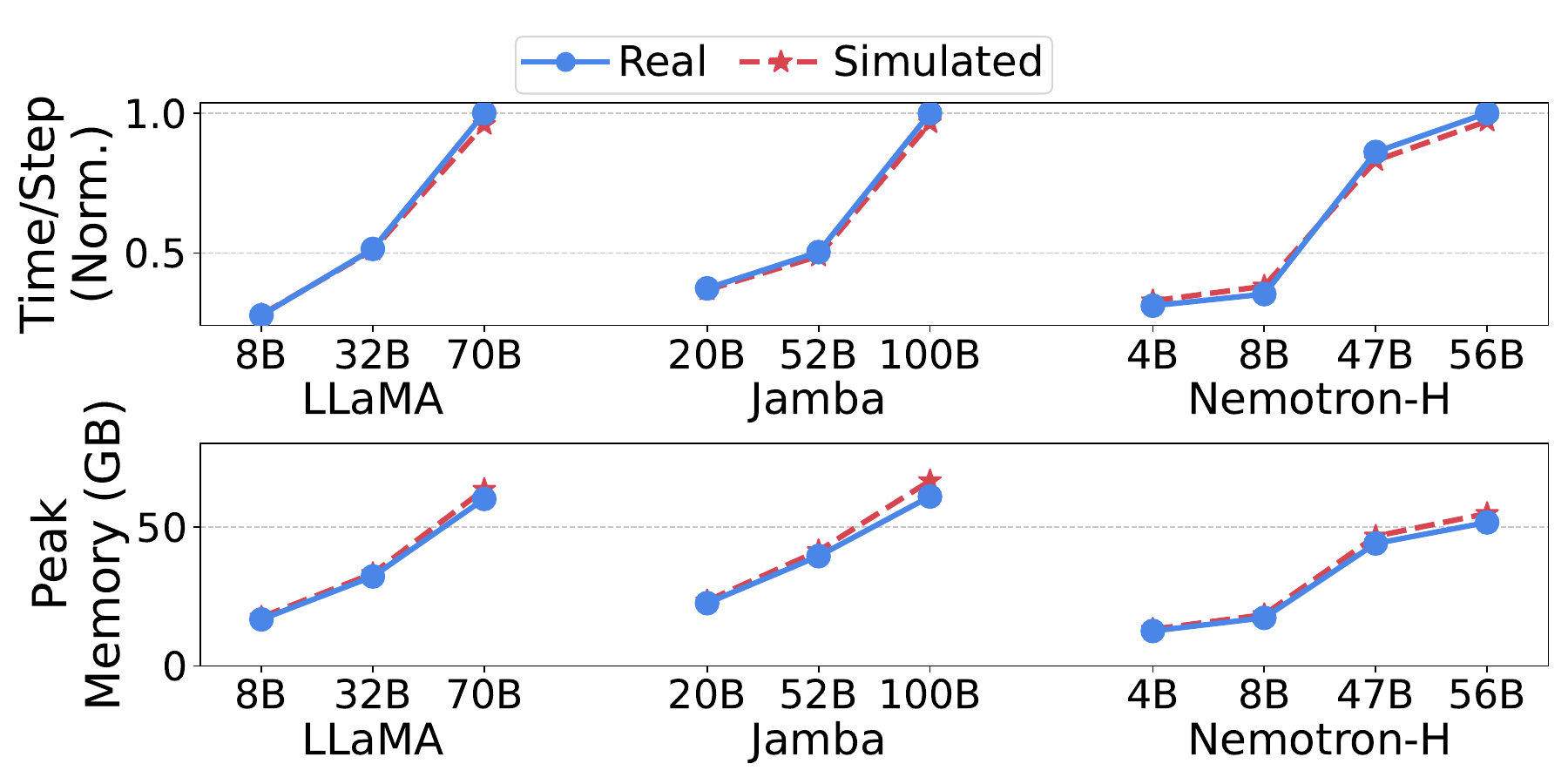}
    \vspace{-20pt}
    \caption{Time and peak memory model fidelity results.}
    \label{fig:sim_thro_mem}
    \vspace{-15pt}
\end{figure}

\subsection{Performance Model Fidelity}\label{sec:sim_acc}
Our simulator demonstrates high fidelity in modeling both execution time and memory consumption. For execution time per step, the relative error ranges from 0.88\% to 4.13\%, with an average of 3.38\%. For memory consumption, the relative error ranges from 5.00\% to 9.24\%, with an average of 5.53\%.
For execution time, the prediction error primarily stems from the gap between idealized simulation and real system execution. 
The simulator assumes deterministic execution times for both computation and communication, while actual runs exhibit performance variability due to hardware-level noise and runtime fluctuations. 
For example, in Nemotron-H 47B, two identical Mamba layers exhibit profiled execution times of $3.73 \pm 0.239$\,ms, leading to small imbalances (approximately 6.4\%) that are not captured by the simulator.
For memory, the simulator slightly overestimates peak memory usage because it relies on theoretical upper bounds to ensure safety against OOM errors. 
Specifically, the simulator does not model certain runtime memory optimizations, such as early release of communication buffers and activation tensors, which reduce memory usage in practice. 
As a result, the predicted peak memory is slightly higher than the measured values.

\subsection{Pipeline Search Cost}\label{sec:pgt}
\begin{figure}[t]
    \centering
    \includegraphics[width=1\linewidth]{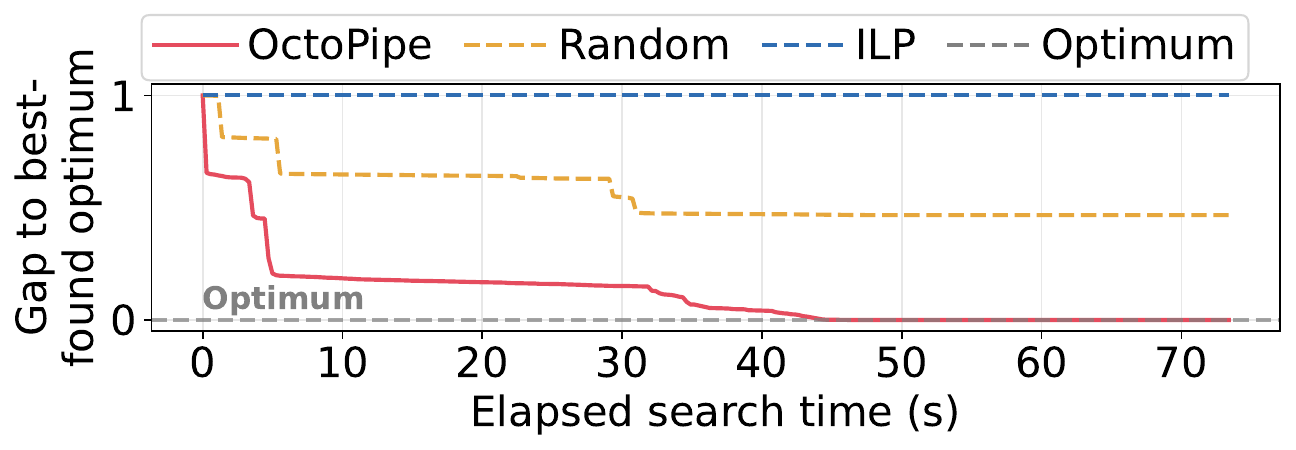}
    \vspace{-10pt}
    \caption{Search processes on a scaled 224-layer variant of Nemotron-H 8B.}    
    \label{fig:convergency_nemotronh}
    \vspace{-15pt}
\end{figure}

\noindent\textbf{Iterative Search Efficiency.}
\autoref{fig:convergency_nemotronh} compares the search processes of \SysName{}, random search, and ILP-based search~\cite{qi2024zerobubble,lin2024tessel} on a scaled Nemotron-H 8B variant whose number of layers is increased from 52 to 224.
\SysName{} adopts iterative bubble-aware tuning (\S\ref{sec:pipeline_tuner}), while random search follows the same iterative framework but selects tuning phases randomly without considering bubble characteristics. 
ILP-based search relies on commercial solvers~\cite{gurobi} to jointly optimize the pipeline schedule.
Due to the combinatorial search space, ILP-based methods (e.g., Tessel~\cite{lin2024tessel}) fail to produce a feasible solution within the same time budget. In contrast, both \SysName{} and random search quickly obtain feasible solutions and continuously refine them through iterative tuning, demonstrating the efficiency and practicality of the iterative search paradigm.

\begin{table}[t]
\caption{Phase-wise breakdown of tuning iterations for \SysName{} and random search. Percentages indicate the allocation of tuning iterations across phases.}
\vspace{-5pt}
\centering
\footnotesize
\setlength{\tabcolsep}{2pt}
\begin{tabular}{
    p{1.25cm}
    *{4}{>{\centering\arraybackslash}p{1.6cm}}
}
\toprule
\textbf{Method} &
\textbf{Total Iter.} &
\textbf{Partitioning} &
\textbf{Placement} &
\textbf{Scheduling} \\
\midrule
OctoPipe & 319 (100\%) & 165 (51.7\%) & 111 (34.8\%) & 43 (13.5\%) \\
Random & 326 (100\%) & 116 (35.6\%) & 103 (31.6\%) & 107 (32.8\%) \\
\bottomrule
\end{tabular}
\vspace{-5pt}
\label{tab:detail_search_overhead}
\end{table}

\noindent\textbf{Bubble-Awareness Analysis.}
\autoref{tab:detail_search_overhead} shows the breakdown of tuning iterations on each phase in \autoref{fig:convergency_nemotronh}.
Without bubble awareness, random search frequently optimizes placement and scheduling even when substantial workload imbalance persists across stages. In such cases, the optimization potential of placement and scheduling is inherently limited, as bubbles are dominated by imbalance.
In contrast, \SysName{} prioritizes tuning partitioning early in the search to mitigate workload imbalance. As shown in \autoref{tab:detail_search_overhead}, \SysName{} allocates 165 iterations to partitioning, whereas random search only allocates 116. This ensures a balanced foundation before the search progressively refines placement to reduce boundary bubbles and scheduling to eliminate residual bubbles. 
This staged, bubble-aware strategy enables \SysName{} to allocate search time more effectively and achieve better pipeline schedules than random search.

\begin{table}[t]
\caption{Strong scaling and weak scaling experimental results measured in Tokens/GPU/Second.}
\vspace{-5pt}
\centering
\setlength{\tabcolsep}{4pt}
\begin{tabular}{llccccc}
\toprule
 & \textbf{\#GPUs} & \textbf{1F1B} & \textbf{Interleaved} & \textbf{Tessel} & \textbf{Mist} & \textbf{OctoPipe} \\
\midrule
\multirow{3}{*}{\makecell{Strong\\scaling}}
 & 32  & 1407 & 1448 & 1445 & 1535 & 1791 \\
 & 64  & 1328 & 1357 & 1368 & 1462 & 1689 \\
 & 128 & 1165 & 1179 & 1199 & 1294 & 1519 \\
\midrule
\multirow{3}{*}{\makecell{Weak\\scaling}}
 & 32  & 1407 & 1448 & 1445 & 1535 & 1791 \\
 & 64  & 1369 & 1419 & 1410 & 1504 & 1739 \\
 & 128 & 1328 & 1360 & 1372 & 1460 & 1688 \\
\bottomrule
\end{tabular}
\label{tab:scaling_res}
\vspace{-5pt}
\end{table}

\subsection{Scalability}
\autoref{tab:scaling_res} presents strong and weak scaling results on Nemotron-H 8B, measured in Tokens/GPU/Second.

\noindent\textbf{Strong Scaling.}
We increase the DP size while keeping the global batch size fixed. Since the micro-batch size remains 1, the number of micro-batches processed by each pipeline replica decreases as the DP size increases. This reduction limits the scheduling opportunities available to \SysName{} and makes bubble reduction more challenging. Nevertheless, \SysName{} consistently outperforms all baselines and achieves its largest relative gain at 128 GPUs, demonstrating that co-optimizing partitioning, placement, and scheduling remains effective at larger scales.

\noindent\textbf{Weak Scaling.}
We increase the DP size and scale the global batch size proportionally, keeping the workload per pipeline replica unchanged. \SysName{} consistently reduces pipeline bubbles through co-optimization and therefore maintains stable per-GPU throughput and a consistent advantage over the baselines as the system scales.

\subsection{Case Study}
\begin{table}[t]
\caption{Large-scale simulation experiments on 8192 GPUs. Throughput results are normalized to 1F1B.}
\vspace{-5pt}
\centering
\label{tab:throughput_results}
\setlength{\tabcolsep}{4pt}
\begin{tabular}
{
    lcccccc
}
\toprule
\textbf{Model} & \textbf{Size} & \textbf{1F1B} & \textbf{Interleaved} & \textbf{Tessel} & \textbf{Mist} & \textbf{OctoPipe} \\ 
\midrule
LLaMA       & 70B & 1.00$\times$ & 1.08$\times$ & 1.11$\times$ & 1.12$\times$ & 1.17$\times$ \\
Jamba    & 100B & 1.00$\times$ & 1.06$\times$ & 1.08$\times$ & 1.11$\times$ & 1.22$\times$ \\
Nemotron-H & 56B & 1.00$\times$ & 0.99$\times$ & 1.09$\times$ & 1.14$\times$ & 1.21$\times$ \\ 
\bottomrule
\end{tabular}
\vspace{-15pt}
\end{table}

\begin{table}[t]
\caption{Throughput results (normalized to 1F1B) with recomputation enabled.}
\vspace{-5pt}
\centering
\setlength{\tabcolsep}{3pt}
\begin{tabular}{lcccccc}
\toprule
\textbf{Model} & \textbf{Size} & \textbf{1F1B} & \textbf{Interleaved} & \textbf{Tessel} & \textbf{Mist} & \textbf{\SysName{}} \\
\midrule
LLaMA        & 70B & 1.00$\times$ & 1.09$\times$ & 1.12$\times$ & 1.11$\times$ & 1.21$\times$ \\
Jamba     & 100B & 1.00$\times$ & 1.09$\times$ & 1.08$\times$ & 1.14$\times$ & 1.34$\times$ \\
Nemotron-H   & 56B & 1.00$\times$ & 1.06$\times$ & 1.09$\times$ & 1.15$\times$ & 1.33$\times$ \\
\bottomrule
\end{tabular}
\label{tab:recompute_throughput}
\vspace{-15pt}
\end{table}
\noindent\textbf{Large-Scale Simulation.}
We evaluate \SysName{} in large-scale settings via simulation due to resource limitations. We scale the DP size and proportionally increase the number of micro-batches, while keeping all other configurations fixed, to emulate training on 8192 GPUs.
\autoref{tab:throughput_results} shows that \SysName{} consistently outperforms baselines, achieving 1.17--1.22$\times$ throughput over 1F1B across models. These results are consistent with the weak scaling trends, indicating that the benefits of \SysName{} persist at large scale.

\noindent\textbf{Incorporating Recomputation.}
We evaluate \SysName{} with recomputation~\cite{chen2016recomp}. As shown in \autoref{tab:recompute_throughput}, \SysName{} continues to outperform 1F1B with recomputation. Because recomputation increases the total computation time, it reduces the fraction of step time attributable to bubbles. Consequently, the speedup of \SysName{} over 1F1B can be lower than without recomputation. Nevertheless, reduced memory pressure allows the tuner to explore more effective scheduling choices within a sufficient memory budget. These results also demonstrate that our simulator accurately captures memory dynamics and effectively guides pipeline schedule tuning.

\section{Related Work}
\noindent\textbf{Data Heterogeneity.}
Variations in sequence lengths lead to fluctuations in micro-batch execution time~\cite{zhang2024disttrain, wang2025flexsp, jiang2024dynapipe, pipeweaver}, resulting in bubbles. To mitigate this issue, prior works adopt dynamic strategies such as micro-batch reordering~\cite{jiang2024dynapipe,zhang2024disttrain}, adaptive parallelism~\cite{wang2025flexsp}, and data rebalancing~\cite{ge2025bytescale}. These approaches primarily target runtime variability induced by input data, rather than heterogeneity within models.

\noindent\textbf{Cross-Model Heterogeneity.}
Multimodal LLMs combine multiple sub-models to process different modalities, introducing cross-model heterogeneity that leads to bubbles. DistTrain~\cite{zhang2024disttrain} and DISTMM~\cite{DISTMM} mitigate this by assigning different parallelism configurations to each sub-model, while DIP~\cite{pipeweaver} distributes layers from each sub-model evenly across stages to balance workloads. Optimus~\cite{Optimus} schedules encoder computation to fill bubbles in multimodal LLM training. However, these approaches focus on cross-model heterogeneity. In contrast, \SysName{} targets intra-model heterogeneity and co-optimizes pipelines to reduce bubbles.

\noindent\textbf{Automatic Parallelization.}
Automatically identifying efficient parallelization strategies~\cite{li2022amp, miao2022galvatron, lai2023merak, liu2024aceso, WLB-LLM} improves training throughput. Recent approaches~\cite{um2024metis, zheng2022alpa, liu2024aceso, zhu2025mist} optimize partitioning by adjusting the number of layers per stage to reduce bubbles. However, these methods do not consider co-optimization to further reduce bubbles.

\noindent\textbf{Memory Optimizations.}
Many pipeline systems~\cite{liu2025mario, zhu2025mist, sun2024adapipe, kim2023bpipe, SlimPipe} adopt \textit{recomputation}~\cite{chen2016recomp} and \textit{offload}~\cite{ren2021zero-offload} to reduce memory overhead. While effective for memory saving, these techniques are largely complementary to \SysName{}.

\section{Conclusion}
We present \SysName{}, a pipeline parallelism system that mitigates pipeline bubbles for heterogeneous models. 
\SysName{} combines a graph-based pipeline simulator for accurate performance estimation with an iterative, bubble-aware tuner that co-optimizes partitioning, placement, and scheduling. We design a unified pipeline executor with deadlock-free, overlap-aware reordering to avoid deadlocks and maximize computation–communication overlap. 
Experiments demonstrate that \SysName{} achieves high simulation accuracy for both execution time and peak memory. It enables efficient pipeline schedule search and improves training throughput compared to state-of-the-art PP systems in our evaluation on various models.

\section*{Acknowledgment}
We are grateful to the anonymous reviewers for their constructive feedback and valuable suggestions. We also thank our colleagues for their insightful discussions and support in developing and evaluating the techniques presented in this paper. This work was supported by the New Generation Artificial Intelligence National Science and Technology Major Project (No.~2025ZD0123805).

\bibliographystyle{IEEEtran}
\bibliography{ref}
\end{document}